\documentclass{PoS}
\usepackage[utf8]{inputenc}
\usepackage{color}
\usepackage{float}
\usepackage{amsmath}
\usepackage{pifont}
\usepackage{graphicx}
\usepackage{subfigure}
\usepackage{dsfont}

\newcommand{\X}{\ding{53}}

\title{The multi-flavor Schwinger model with chemical potential - Overcoming the sign problem with Matrix Product States}

\ShortTitle{Multi-flavor QED$_2$ with chemical potential - Overcoming the sign problem with MPS}

\author{Mari Carmen Ba\~nuls\\
Max-Planck-Institut f\"ur Quantenoptik, Hans-Kopfermann-Stra{\ss}e 1, 85748 Garching\\
E-mail: \email{banulsm@mpq.mpg.de}}
\author{Krzysztof Cichy\\
Goethe-Universit\"at Frankfurt am Main, Institut f\"ur Theoretische Physik,
Max-von-Laue-Stra{\ss}e 1, 60438 Frankfurt am Main, Germany\\
Faculty of Physics, Adam Mickiewicz University, Umultowska 85, 61-614 Pozna\'{n}, Poland\\
E-mail: \email{kcichy@th.physik.uni-frankfurt.de}}
\author{J. Ignacio Cirac\\
Max-Planck-Institut f\"ur Quantenoptik, Hans-Kopfermann-Stra{\ss}e 1, 85748 Garching, Germany\\
E-mail: \email{ignacio.cirac@mpq.mpg.de}}
\author{Karl Jansen\\
NIC, DESY Zeuthen, Platanenallee 6, 15738 Zeuthen, Germany\\
E-mail: \email{Karl.Jansen@desy.de}}
\author{\speaker{Stefan K\"uhn}\\
Max-Planck-Institut f\"ur Quantenoptik, Hans-Kopfermann-Stra{\ss}e 1, 85748 Garching, Germany\\
E-mail: \email{stefan.kuehn@mpq.mpg.de}}
\author{Hana Saito\\
AISIN AW Co., Ltd.\\
10 TAKENE, FUKII-CHO, ANJO,\\
AICHI, 444-1192, JAPAN}

\abstract{During recent years there has been an increasing interest in the application of matrix product states, and more generally tensor networks, to lattice gauge theories. This non-perturbative method is sign problem free and has already been successfully used to compute mass spectra, thermal states and phase diagrams, as well as real-time dynamics for Abelian and non-Abelian gauge models. In previous work we showed the suitability of the method to explore the zero-temperature phase structure of the multi-flavor Schwinger model at non-zero chemical potential, a regime where the conventional Monte Carlo approach suffers from the sign problem. Here we extend our numerical study by looking at the spatially resolved chiral condensate in the massless case. We recover spatial oscillations, similar to the theoretical predictions for the single-flavor case, with a chemical potential dependent frequency and an amplitude approximately given by the homogeneous zero density condensate value.}

\FullConference{34th annual International Symposium on Lattice Field Theory\\
         24-30 July 2016\\
         University of Southampton, UK}

\begin{document}

\section{Introduction}
A lot of our current understanding for many gauge models has been gained thanks to lattice gauge theory (LGT). Formulating gauge theories on a discretized space-time lattice with Euclidean time allows the application of powerful Markov chain based Monte Carlo methods that make it possible to explore mass spectra, phase diagrams and many other phenomena numerically. However, the infamous sign problem~\cite{Troyer2005} is a major obstacle for the numerical exploration of lattice gauge theories in certain parameter regimes, as the appearance of negative or even complex probability amplitudes prevents efficient Monte Carlo sampling. As a consequence, there are questions that cannot be addressed with this method, such as large parts of the phase diagram of quantum chromodynamics (QCD) at non-zero temperature and baryon density. Hence, there is great interest in alternative numerical methods overcoming the sign problem~\cite{Gattringer2016,Cristoforetti2013,Ammon2016}. In the last years, methods based on tensor networks, such as matrix product states (MPS), have revealed themselves as valid candidates to explore the Hamiltonian formulation of LGT. Tensor networks, originally developed in the realm of quantum information theory, do not suffer from the sign problem and during the last years they have already been successfully applied to LGT problems and have demonstrated their power for mass spectra~\cite{Banuls2013,Buyens2013,Buyens2015}, thermal states~\cite{Banuls2015,Buyens2016,Banuls2016} and phase diagrams~\cite{Silvi2016} as well as for simulating dynamical problems for both Abelian and non-Abelian gauge theories~\cite{Buyens2013,Kuehn2015}. Recently, we used MPS to compute the phase structure of the two-flavor Schwinger model with non-zero chemical potential~\cite{Banuls2016a} and performed a lattice calculation, with full extrapolation procedure reaching the proper continuum limit, in a regime where the sign problem occurs.

Here we extend our study and address the question of spatial inhomogeneities in the chiral condensate for the two-flavor Schwinger model at non-zero chemical potential. Taking advantage of the fact that the MPS approach to the Hamiltonian formulation of LGT is not only free of the sign problem but also explicitly yields the ground state wave function, we can compute the spatially resolved chiral condensate. We find that the  condensate shows sinusoidal oscillations with a frequency that depends linearly on the isospin density, and an amplitude close to the (spatially homogeneous) zero density condensate value. The observed behavior is similar to the theoretical predictions for the single-flavor case ~\cite{Fischler1979,Kao1994,Christiansen1996} and consistent with analytical calculations for fermion bilinears in the two-flavor case~\cite{Christiansen1997}.

\section{Model \& Methods}
We consider the discrete Hamiltonian formulation of the multi-flavor Schwinger model with Kogut-Susskind staggered fermions~\cite{Kogut1975}. The dimensionless Hamiltonian for $F$ flavors on a finite lattice with open boundary conditions (OBC) consisting of $N$ sites with a lattice spacing $a$ is given by
 \begin{align*}
 W = &-ix\sum_{n=0}^{N-2}\sum_{f=0}^{F-1}\left(\phi^\dagger_{n,f}e^{i\theta_n}\phi_{n+1,f}-\mathrm{h.c.}\right)+\sum_{n=0}^{N-1}\sum_{f=0}^{F-1}\left(\mu_f(-1)^n +\nu_f \right)\phi^\dagger_{n,f}\phi_{n,f}+ \sum_{n=0}^{N-2} L_n^2,
\end{align*}
where $\phi_{n,f}$ is a single component fermionic field for flavor $f$ on site $n$. The operators $L_n$ and $\theta_n$ act on the links between sites $n$ and $n+1$ and fulfill the commutation relation $\left[\theta_n,L_m\right] = i\delta_{n,m}$. Hence they are canonical conjugates and $L_n$ gives the quantized electric flux on link $n$, whereas $e^{i\theta_n}$ acts as raising operator. We work with a compact formulation where $\theta_n$ is restricted to $[0,2\pi]$. The dimensionless parameters in the Hamiltonian are $x = 1/(ag)^2$, $\mu_f = 2\sqrt{x}m_f/g$ and $\nu_f = 2\sqrt{x}\kappa_f/g$, where $m_f$ ($\kappa_f$) is the bare fermion mass (chemical potential) for flavor $f$ and we use the coupling $g$ to set the scale. Physical states have to fulfill the Gauss law, $ L_n - L_{n-1} = \sum_{f=0}^{F-1} \left(\phi_{n,f}^\dagger\phi_{n,f}-\frac{1}{2}\left(1-(-1)^n\right)\right)$. In one spatial dimension, and OBC, we can integrate out the gauge degrees of freedom~\cite{Hamer1997}. We assume the electric field to be zero on the left boundary. After applying a residual gauge transformation, we obtain a formulation restricted to the gauge invariant subspace:
\begin{align}
\begin{aligned}
 W =& -ix\sum_{n=0}^{N-2}\sum_{f=0}^{F-1}\left(\phi^\dagger_{n,f}\phi_{n+1,f}-\mathrm{h.c.}\right)
 +\sum_{n=0}^{N-1}\sum_{f=0}^{F-1}\left(\mu_f(-1)^n +\nu_f \right)\phi^\dagger_{n,f}\phi_{n,f}  \\
 &+ \sum_{n=0}^{N-2} \left( \sum_{k=0}^n\left(\sum_{f=0}^{F-1}\phi_{k,f}^\dagger\phi_{k,f}-\frac{F}{2}(1-(-1)^k)\right)\right)^2.
\end{aligned}
\label{hamiltonian_dimensionless}
\end{align}
In the following we focus on the massless case for two flavors, for which the model has an SU(2) flavor symmetry, and we target the subspace of vanishing total charge. Moreover, it can be shown that the Hamiltonian (\ref{hamiltonian_dimensionless}) only depends on the difference between the chemical potentials up to a constant~\cite{Banuls2016a}. Thus we fix $\nu_0=0$ and only vary $\nu_1$ to introduce an imbalance between the two flavors leading to a non-vanishing isospin number $\Delta N = N_0-N_1$, $N_i=\sum_{n=0}^{N-1}\phi^\dagger_{n,i}\phi_{n,i}$, and entering a region where the conventional Monte Carlo approach suffers from the sign problem. 

We are interested in the effect of this imbalance on the chiral condensate in the ground state. In the large $N_c$ limit of QCD at high density, the chiral condensate shows spatial oscillations~\cite{Deryagin1992}. Also for the single-flavor Schwinger model, which is in many aspects similar to QCD, analytical computations show that the condensate at any finite density has the form of a standing wave~\cite{Fischler1979,Kao1994,Christiansen1996}, $\langle \bar{\psi}(y)\psi(y)\rangle = \langle\bar{\psi}\psi\rangle_0\cos(2\kappa y)$, where $\psi$ is a two component Dirac spinor, $\langle\bar{\psi}\psi\rangle_0$ the (spatially homogeneous) zero density expectation value for the chiral condensate and $\kappa$ is the chemical potential. A possible explanation for the oscillations, put forward in Ref.~\cite{Metlitski2007}, linked the phenomenon to the breaking of translational invariance due to the introduction of a uniform background charge. For the multi-flavor case, which was addressed in Ref.~\cite{Christiansen1997}, it was also found that fermion bilinears show spatial inhomogeneities. 

In order to address this question, we variationally compute the ground state of the Hamiltonian (\ref{hamiltonian_dimensionless}) using MPS. The MPS ansatz for a system of $N$ sites with OBC is given by
\begin{align*}
 |\psi\rangle = \sum_{i_0,i_1,\dots i_{N-1}} M^{i_0}_0M^{i_1}_1 \dots M^{i_{N-1}}_{N-1}|i_0\rangle \otimes \dots\otimes |i_{N-1}\rangle,
\end{align*}
where $M^{i_k}_k$ are complex matrices in $\mathds{C}^{D\times D}$ for $0<k<N-1$, $M^{i_0}_0$ ($M^{i_{N-1}}_{N-1}$) is a row (column) vector and $|i_k\rangle_{k=0}^{d-1}$ is a local basis for the $d$-dimensional Hilbert space on site $k$. The parameter $D$, the bond dimension of the MPS, determines the number of variational parameters and limits the amount of entanglement that can be present in the state (a detailed review on MPS methods can be found in Ref.~\cite{Verstraete2008}). Expectation values of local observables can be efficiently computed for a MPS, so that we can find the spatially resolved condensate in the ground state. In our lattice formulation this corresponds to $C(y=2n/\sqrt{x})=\sum_{f=0}^{1}(C_{n,f} + C_{n+1,f})$, $n=0,2,4,6,\dots$, where $C_{n,f} = \frac{\sqrt{x}}{N} (-1)^n\phi^\dagger_{n,f}\phi_{n,f}$. The condensate is summed over each pair of even and odd neighboring sites to account for the staggered formalism\footnote{For convenience in the visualization, we have also summed both flavors for each site. However, we observe the same behavior for each individual flavor.}. We focus on fixed volumes $Lg=N/\sqrt{x}$ ranging from $2$ to $16$, where $L$ is the physical volume of the system. For our analysis, we use a very fine lattice spacing, corresponding to $x=1024$, for which lattice effects are very small. Moreover, we fix $D=160$ for all the rest which is sufficiently large to control truncation errors well enough.

\section{Results}
\begin{figure}
 \centering
 \subfigure[Phase structure]{\label{fig:t1_phase_structure}
\includegraphics[width=0.49\textwidth]{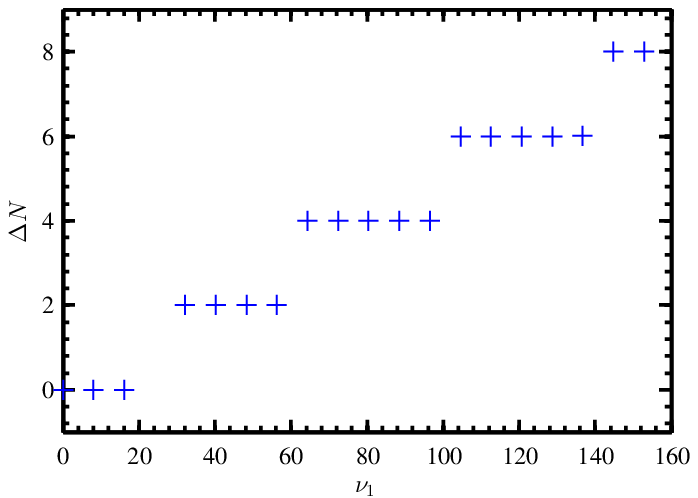}}
 \subfigure[Site resolved condensate for different phases]{\label{fig:t1_c_osci}
\includegraphics[width=0.49\textwidth]{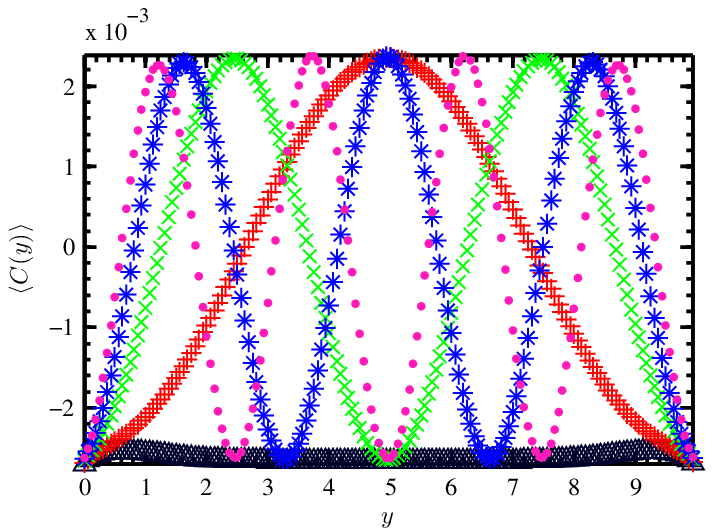}}
\caption{Left: Phase structure of the model as a function of the chemical potential for $Lg=10$. Right: Expectation value of the chiral condensate as a function of position for $Lg=10$. The different curves correspond to different phases characterized by different $\Delta N$, black triangles represent $\Delta N=0$, red crosses $\Delta N=2$, green \X's $\Delta N=4$, blue asterisks $\Delta N=6$ and magenta dots $\Delta N=8$.}
\label{fig:t1}
\end{figure}

In Ref.~\cite{Banuls2016a} we studied the phase structure of the model at zero temperature with a similar setup. We found that the different phases are characterized by the isospin number, which does not change continuously as $\nu_1$ is increased but instead shows abrupt jumps, corresponding to first-order phase transitions, as can be seen in Fig.~\ref{fig:t1_phase_structure} for $Lg=10$. To assess the effect of the chemical potential (and hence different isospin number) on the condensate, we choose a point in every phase and compute the the position dependent condensate, $\langle C(y) \rangle$, for each case. The results are shown in Fig.~\ref{fig:t1_c_osci}. For $\Delta N=0$, the condensate is homogeneous, except close to the edges, what we interpret as small finite size effects. For $\Delta N>0$, by contrast, we observe sinusoidal oscillations with an amplitude close to the condensate value at vanishing $\Delta N$. At a fixed volume, the frequency of the oscillations increases with the isospin number, $\Delta N$, and we observe $\Delta N/2$ oscillation periods inside our system.

\begin{figure}
 \centering
 \subfigure[Frequency]{\label{fig:t2_freq}
\includegraphics[width=0.49\textwidth]{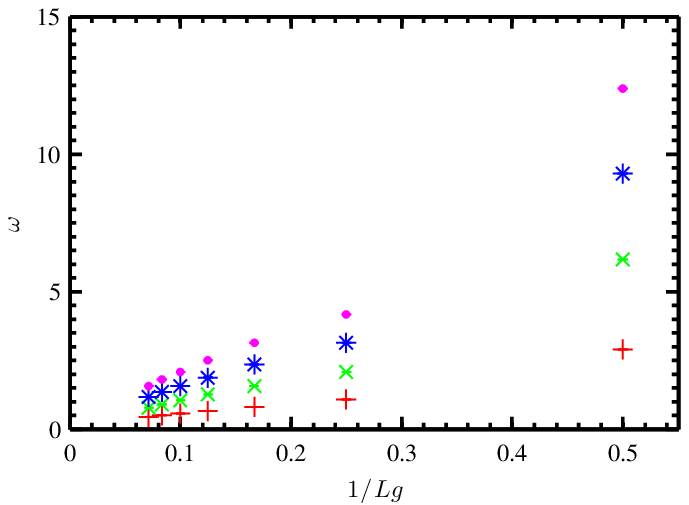}}
 \subfigure[Amplitude]{\label{fig:t2_amp}
\includegraphics[width=0.49\textwidth]{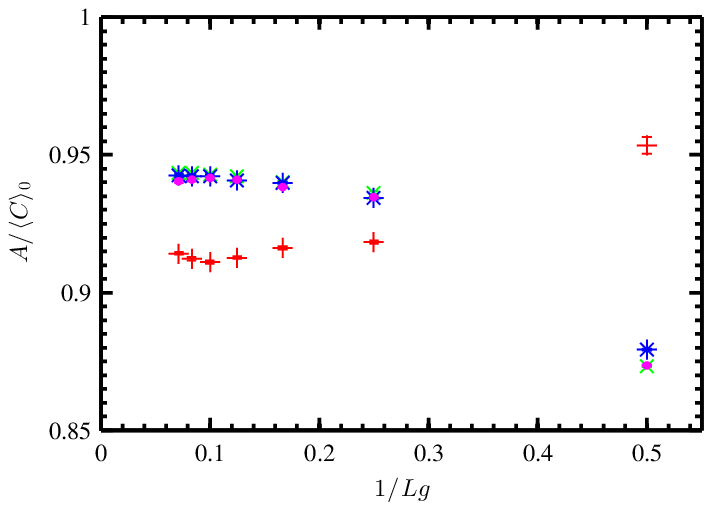}}
\caption{Frequency $\omega$ (left) and amplitude $A$ in units of the zero density expectation value (right) as a function of inverse volume for different phases, where the red crosses represent $\Delta N=2$, the green \X's $\Delta N=4$, the blue asterisks $\Delta N=6$ and the magenta dots $\Delta N=8$.}
\label{fig:t2}
\end{figure}

\begin{figure}
 \centering
 \subfigure[Phase shift]{\label{fig:t3_theta}
\includegraphics[width=0.49\textwidth]{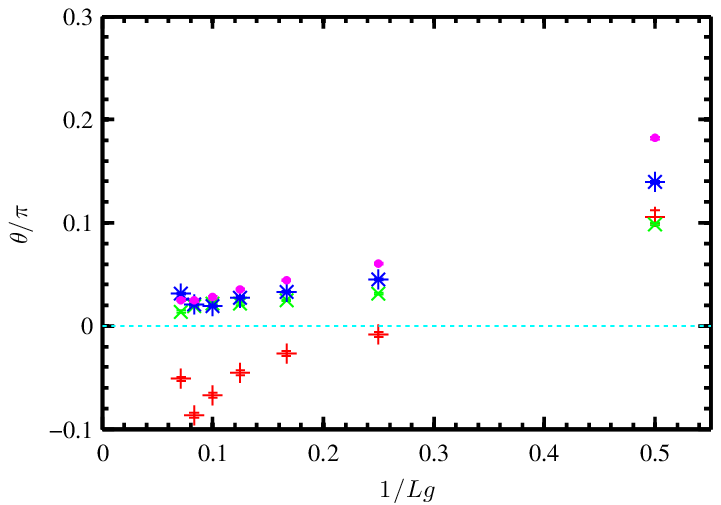}}
 \subfigure[Offset]{\label{fig:t3_offset}
\includegraphics[width=0.49\textwidth]{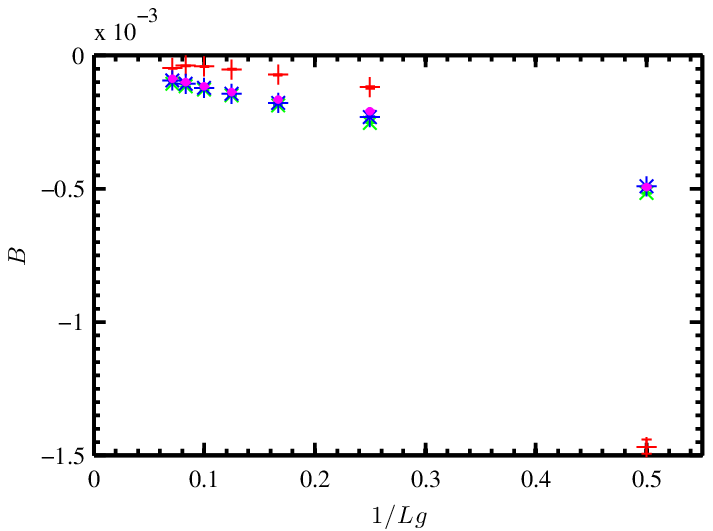}}
\caption{Phase shift $\theta$ (left) and offset $B$ (right) as a function of inverse volume for different phases, where the red crosses represent $\Delta N=2$, the green \X's $\Delta N=4$, the blue asterisks $\Delta N=6$ and the magenta dots $\Delta N=8$.}
\label{fig:t3}
\end{figure}

Our data suggest that the oscillations for $\Delta N>0$ are of the form $\langle C(y)\rangle = A\cos\left(\omega y + \theta \right)+ B$ and thus we fit our data to this function to extract the amplitude $A$, frequency $\omega$, phase shift $\theta$ and offset $B$. Determining these parameters for several system sizes and several phases, we obtain the results shown in Figs.~\ref{fig:t2} and \ref{fig:t3}, where the error bars correspond to $1\sigma$ confidence intervals for the fit parameters\footnote{Notice that we have not performed a continuum extrapolation (different to Ref.~\cite{Banuls2016a}), but the results presented correspond to a fixed (albeit very small) lattice spacing.}. The frequency, depicted in Fig.~\ref{fig:t2_freq}, shows a clear dependence on the isospin density, $\Delta N/Lg$. For $Lg>2$ we see a linear decrease with increasing volume for each phase, as inside each phase we have $\Delta N/2$ oscillation periods, independently of volume. The deviations for volume $2$ are likely due to remaining lattice effects, which become increasingly important for small volumes and large $\Delta N$, as observed in Ref.~\cite{Banuls2016a}. As Fig.~\ref{fig:t2_amp} reveals, the amplitudes of the oscillations are close to the expectation value of the condensate in the $\Delta N=0$ phase, $\langle C \rangle_0$. We observe that for $\Delta N>2$, the values obtained for $A$ deviate less from $\langle C \rangle_0$ than those for $\Delta N=2$ and that there is hardly any change with volume, except for the smallest volume $Lg=2$.

The phase shift, $\theta$, depicted in Fig.~\ref{fig:t3_theta}, shows a similar behavior. While for $\Delta N > 2$ the data approaches zero for increasing volume, there is no clear trend towards zero for $\Delta N=2$. Finally, the offset, $B$, converges to zero for large volumes, with the smallest values found for $\Delta N=2$. For $\Delta N>2$ there is hardly any difference between the offset for different phases anymore.

\section{Discussion \& Outlook}
In summary, we have used MPS to explore the spatial dependence of the ground state chiral condensate in the two-flavor Schwinger model in a regime where the conventional Monte Carlo approach would suffer from the sign problem. We observe a standing wave structure similar to the theoretically predicted behavior for the single-flavor case~\cite{Fischler1979,Kao1994,Christiansen1996} and consistent with analytical calculations for the multi-flavor case~\cite{Christiansen1997}. We have fitted the results to a cosine function and analyzed the behavior of the various parameters in terms of isospin number and volume. For fixed volume, we find that the frequency of the oscillations increases when the isospin number, which characterizes the phase, increases. Inside a phase with a given isospin number $\Delta N$, we see $\Delta N/2$ oscillation periods of the condensate over the spatial extension of the system. Consequently, the frequency shows a linear decrease with the isospin density, $\Delta N/Lg$, except for the smallest volumes considered, which presumably suffer from enhanced lattice effects. The amplitudes of the oscillations are approximately given by the expectation value of the homogeneous condensate at vanishing isospin number, independently of volume. The observed offsets show a clear tendency towards zero for increasing volume, whereas for the phase shifts this tendency is less clearly visible. For phases characterized by $\Delta N>2$ we observe that the shift is getting closer to zero, however for $\Delta N=2$ there is no clear trend recognizable. 

Moreover, it is straightforward to improve the precision by studying additional lattice spacings and bond dimensions, which would allow us to estimate the truncation errors and an extrapolation to the continuum limit.

\section*{Acknowledgments}
K.C.\ was supported by the Deutsche Forschungsgemeinschaft (DFG), project nr. CI 236/1-1 (Sachbeihilfe).

\bibliographystyle{JHEP}
\bibliography{Papers_POS_converted}
\end{document}